%% file: apssamp.tex
\documentclass[reprint, amsmath, amssymb, aps, prb, onecolumn, nofootinbib]{revtex4-2}

\usepackage{graphicx}
\usepackage{import}
\usepackage{longtable}
\usepackage{booktabs}
\usepackage{dcolumn}
\usepackage{bm}
\usepackage[backref=none]{hyperref} 
\usepackage[usenames,dvipsnames]{color}
\definecolor{MyDarkBlue}{rgb}{0,0.1,0.75}
\hypersetup{pdfborder={0 0 0},colorlinks,breaklinks=true,
  urlcolor={MyDarkBlue},citecolor={MyDarkBlue},linkcolor={MyDarkBlue}
}
\begin{document}

\preprint{APS/123-QED}

\title{A re-examination of antiferroelectric PbZrO$_3$ and PbHfO$_3$: an 80-atom $Pnam$ structure}

\author{J. S. Baker$^{1, 2}$}
\email{jack.baker.16@ucl.ac.uk}
\author{M. Paściak$^3$}
\email{pasciak@fzu.cz}
\author{J. K. Shenton$^4$}
\author{P. Vales-Castro$^5$}
\author{B. Xu$^6$}
\author{J. Hlinka$^3$}
\author{P. Márton$^3$}
\author{R. G. Burkovsky$^7$}
\author{G. Catalan$^{5, 8}$}
\author{A. M. Glazer$^9$}
\author{D. R. Bowler$^{1, 2, 10}$ \vspace{0.2cm}}

\affiliation{$^1$London Centre for Nanotechnology, UCL, 17-19 Gordon St, London WC1H 0AH, UK}

\affiliation{$^2$Department of Physics \& Astronomy, UCL, Gower St, London WC1E 6BT, UK}

\affiliation{$^3$Institute of Physics, Academy of Sciences of the Czech Republic, \ Na Slovance 1999/2, 182 21 Praha 8, Czech Republic}

\affiliation{$^4$Department of Materials, ETH Zurich, CH-8093 Zürich, Switzerland}

\affiliation{$^5$Catalan Institute of Nanoscience and Nanotechnology (ICN2), Campus Universitat Autonoma de Barcelona, Bellaterra 08193, Spain}

\affiliation{$^6$School of Physical Science and Technology, Soochow University, Suzhou 215006, China}

\affiliation{$^7$Peter the Great Saint-Petersburg Polytechnic University, Saint-Petersburg, Russian Federation}

\affiliation{$^8$Institut Català de Recerca i Estudis Avançats (ICREA), Barcelona 08010, Catalunya}

\affiliation{$^9$Department of Physics, University of Oxford, Parks Road, Oxford OX1 3PU, UK}

\affiliation{$^{10}$International Centre for Materials Nanoarchitectonics (MANA) National Institute for Materials Science (NIMS), 1-1 Namiki, Tsukuba, Ibaraki 305-0044, Japan}

\date{\today}

\begin{abstract}
First principles density functional theory (DFT) simulations of antiferroelectric (AFE) PbZrO$_3$ and PbHfO$_3$ reveal a dynamical instability in the phonon spectra of their purported low temperature $Pbam$ ground states. This instability doubles the $c$-axis of $Pbam$ and condenses five new small amplitude phonon modes giving rise to an 80-atom $Pnam$ structure. Compared with $Pbam$, the stability of this structure is slightly enhanced and highly reproducible as demonstrated through using different DFT codes and different treatments of electronic exchange \& correlation interactions. This suggests that $Pnam$ is a new candidate for the low temperature ground state of both materials. With this finding, we bring parity between the AFE archetypes and recent observations of a \textit{very similar} AFE phase in doped or electrostatically engineered BiFeO$_3$.
\end{abstract}

\maketitle

Nearly seventy years have passed since the theory of the antiferroelectric (AFE) phenomenon was proposed by Kittel \cite{Kittel1951} and the observation in PbZrO$_3$ (PZO) by Shirane, Sawaguchi and Takagi \cite{Shirane1951_PZO}. Today, although most consider PZO and PbHfO$_3$ (PHO, PZO's isoelectronic and isostructural partner) the AFE archetypes, many fundamental aspects of these materials remain hotly contested. Indeed, recent years have brought the very nature of the AFE phase transition into question with no clear consensus in sight \cite{Tagantsev2013, Rabe2013, iniguez2014, Fthenakis2017, ValesCastro2018, Xu2019}. Even the crystal structure of the low temperature AFE phase is a point of order. The majority of the community regard the structure as being best described with $Pbam$ symmetry \cite{Fujishita1982, Glazer1993, Corker1997, Madigout1999, Fujishita2003}, but the path to this agreement was a contentious one. Several different space group assignments were proposed (which are summarized in \cite{Fujishita1982}) as well as suggestions of structural disorder \cite{Glazer1993}. Compounded with this, the presence of complex twinning \cite{Scott1972, Fesenko1976, Glazer1993} and incommensurations \cite{Burkovsky2017, Bosak2020} are ubiquitous in these materials making the task of accurate structure refinement a challenging one. Recently, first principles calculations were used to show that a large number of unique and increasingly incommensurate dynamical instabilities were present in the phonon dispersion relations of cubic PZO and ordered models of near-morphotropic PbZr$_x$Ti$_{1 - x}$O$_3$ (PZT, $x=0.5$) \cite{Baker2019}. We then ask: do any further modes of this type condense within the low temperature ground state of PZO and PHO? If yes, what does this mean for the dynamical stability of the AFE $Pbam$ phase? Now that the technological importance of AFE materials have been realized, it is crucial to return to address these fundamental issues to ensure that new AFE technologies including solid state cooling (exploitative of the large negative electrocaloric effect \cite{Bai2011, Geng2015, Guo2019, Valescastro2020}) and energy storage devices \cite{Liu2018} are built upon stable foundations. \par

In this work, we use simulations based on density functional theory (DFT) to show that the phonon dispersion relations of $Pbam$ PZO and PHO feature a single dynamical instability at the Z-point (of $Pbam$), $\mathbf{q}_Z = (0, 0, \frac{1}{2})$. We reason that such instabilities \textit{should not} persist within the low temperature ground state as they would condense into the structure as a soft mode. Following the eigendisplacements of this instability, we demonstrate the emergence of an 80-atom $Pnam$ phase\footnote[1]{We choose to adopt a non-standard representation $Pnam$ space group in place of (the equivalent) $Pnma$ for two reasons. Firstly, using $Pnam$ we retain the same orientation of axes as the well known $Pbam$ structure. This way it is clear that the b-glide (of $P\underline{b}am$) is replaced by an n-glide operation in $P\underline{n}am$. Secondly, using $Pnam$ avoids confusion with the prototype CaTiO$_3$-like $Pnma$ perovskite structure, which is \textit{not similar} to the AFE $Pnam$ phase.} slightly lower in energy than $Pbam$ and described by \textit{eleven distinct phonon modes}. \par

\begin{figure}[hp]
    \centering
    \vspace{5cm}
    \includegraphics{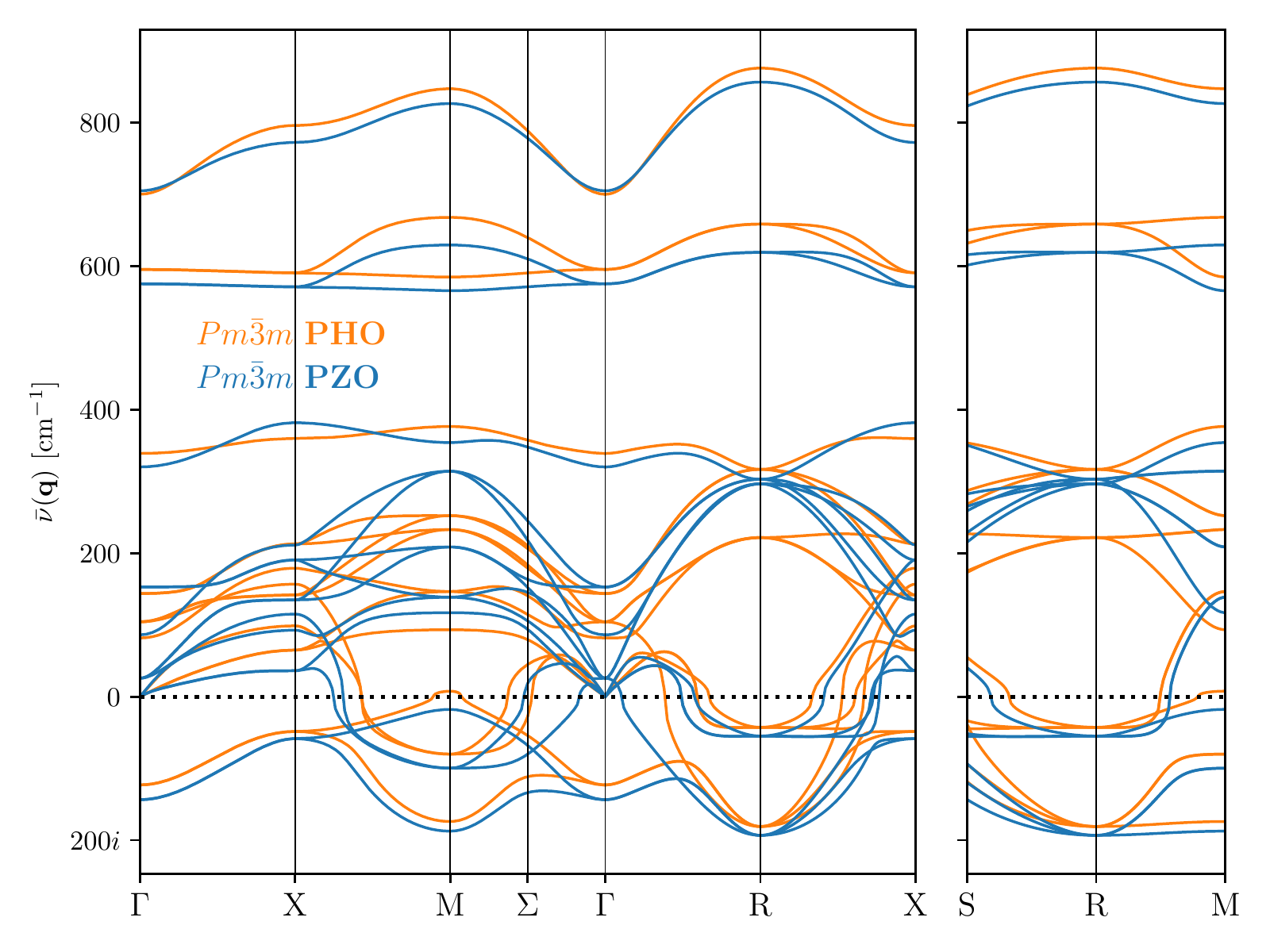}
    \caption{The phonon dispersions of $Pm\bar{3}m$ PZO (blue) and PHO (orange).}
    \label{fig:Pm3mPZOandPHO_phon}
\end{figure}

\begin{figure}[hp]
    \centering
    \vspace{1cm}
    \includegraphics{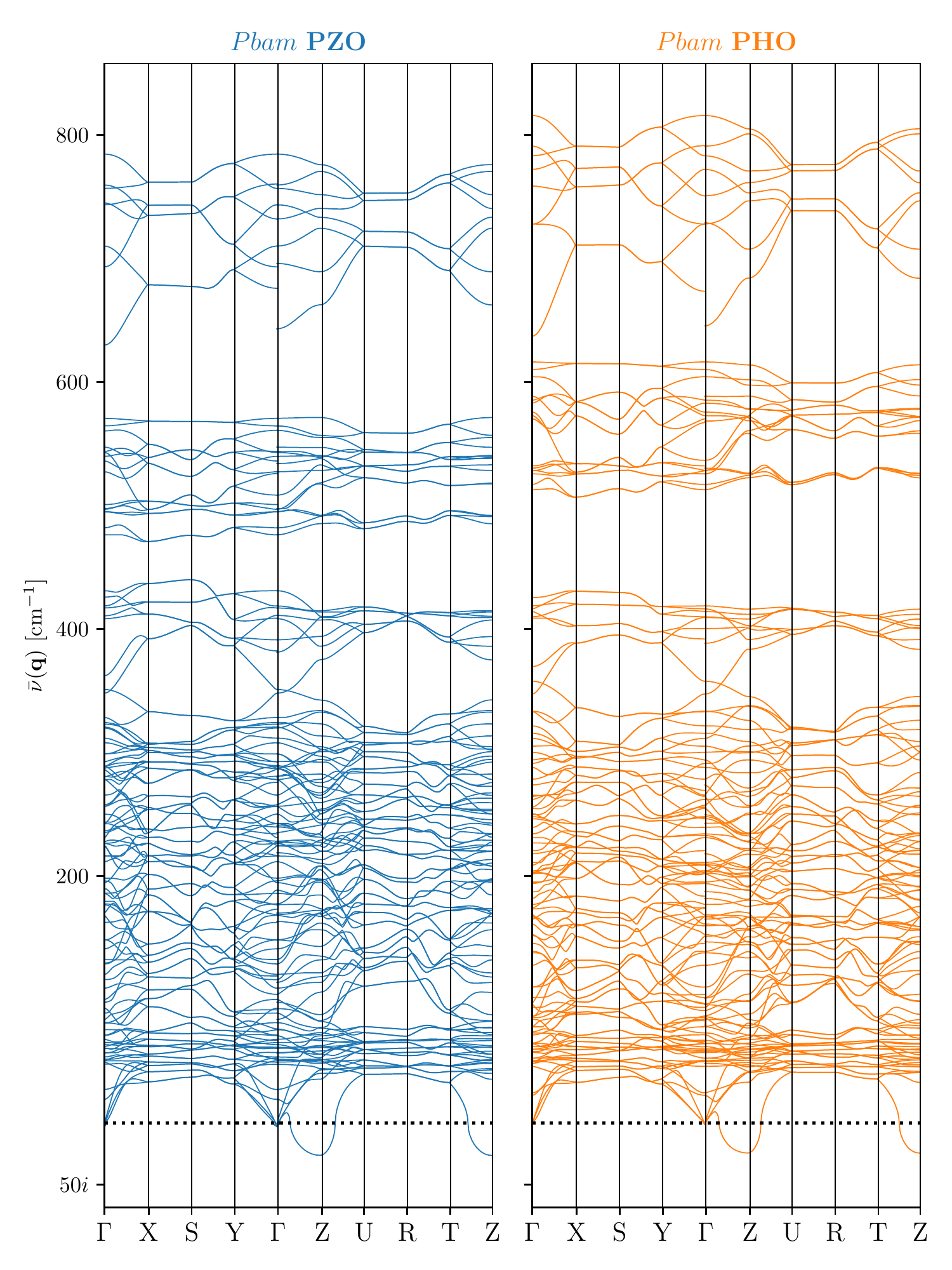}
    \caption{The phonon dispersions of $Pbam$ PZO (blue) and PHO (orange).}
    \label{fig:PbamPZOandPHO_phon}
\end{figure}

The phonon dispersion relations of $Pm\bar{3}m$ (Figure \ref{fig:Pm3mPZOandPHO_phon}) and $Pbam$ (Figure \ref{fig:PbamPZOandPHO_phon}) PZO/PHO are calculated using the implementation of density functional perturbation theory \cite{Gonze1995, Gonze1997, Baroni2001} (DFPT) implemented within \texttt{ABINIT} \cite{Gonze2009, Gonze2016} (\texttt{v8.10.2}). These calculations use the local density approximation of Perdew \& Wang \cite{Perdew1992} (LDA-PW) and projector augmented wave \cite{Blochl1994} (PAW) data sets from the \texttt{JTH} library \cite{Jollet2014} (\texttt{v1.1}). To correct for the undefined nature of long-range coulomb interactions as we approach the $\Gamma$-point \cite{Henry1965}, we apply the non-analytical correction \cite{Gonze1997} to the dispersions, correctly accounting for longitudinal-optical transverse-optical splitting \cite{Zhong1994}. To achieve high accuracy dispersions, we use a 680.29 eV plane wave cutoff, relax the ionic positions to a stringent force tolerance of $1 \times 10^{-6}$ eV/$\text{\AA}$ and interpolate dispersions from exact frequencies calculated on dense $\Gamma$-centered $6 \times 6 \times 6$ (for $Pm\bar{3}m$) and $5 \times 3 \times 3$ (for $Pbam$) $\mathbf{q}$-point meshes. These meshes share dimensions and centering with the Monkhorst-Pack \cite{Monkhorst1976} (MP) $\mathbf{k}$-point meshes used for Brillouin zone (BZ) integrals. \par

While the phonon dispersion of $Pm\bar{3}m$ PZO (Figure \ref{fig:Pm3mPZOandPHO_phon}, blue lines) has been discussed elsewhere \cite{Ghosez1999, Baker2019}, we briefly remark on some important features and compare with the $Pm\bar{3}$m PHO dispersion (Figure \ref{fig:Pm3mPZOandPHO_phon}, orange lines). With a focus on the imaginary branches (indicative of dynamical instabilities) we see that both materials are firmly unstable throughout the entirety of the first BZ. In turn, this gives rise to a large number of unique unstable modes which are discussed in \cite{Baker2019}. We can see clearly the strong instabilities at the R and $\Sigma$-points known to comprise the \textit{majority} of the distortion defining the $Pbam$ AFE phase \cite{Hlinka2014, iniguez2014}. We also note that both materials feature exceptionally flat bands, especially in the vicinity of the R-point and along the most unstable R $\rightarrow$ M lines; a tell-tale sign that a material is prone to structural incommensurations. Comparing the two materials, we see that their imaginary spaces are strongly similar albeit PZO is \textit{slightly} more unstable than PHO. Indeed, when analysing modes at the $\Gamma$, X, M $\Sigma$ and R-points, the character of the instabilities of the two materials are identical (although, this comparison is imperfect for the M$_2^+$ mode as it has a real frequency in PHO). 

\begin{figure}[hp]
    \centering
    \vspace{5cm}
    \includegraphics[width=0.8\linewidth]{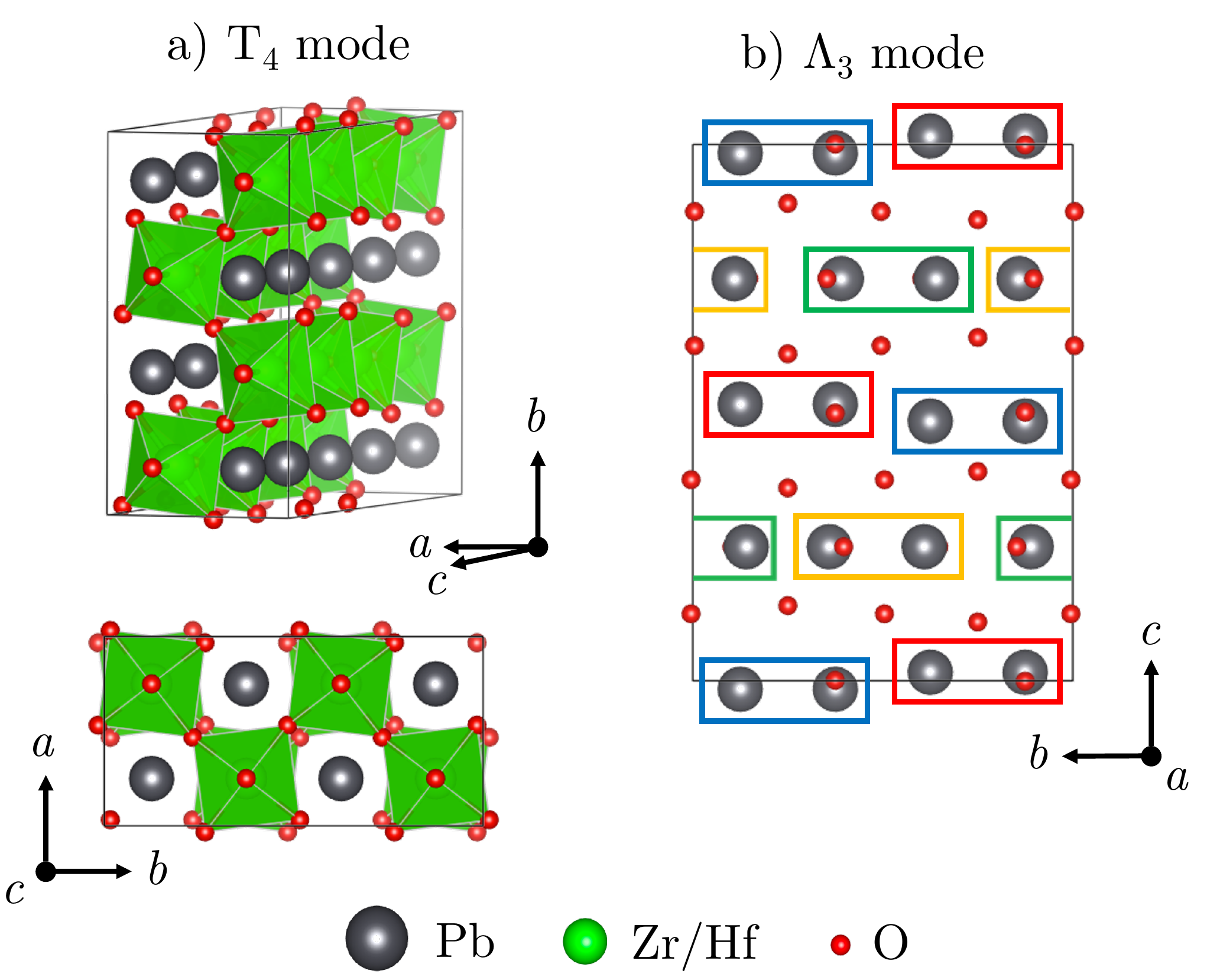}
    \caption{The two most important modes defining the difference between $Pbam$ and $Pnam$ PZO/PHO. The magnitude of the displacement is exaggerated. a) The T$_4$ mode. ZrO$_6$/HfO$_6$ octahedra rotate in antiphase about the $c$ axis in \textit{pairs} ($++$ $--$ rotations). b) The $\Lambda_3$ mode. Antipolar Pb displacements are grouped by a common displacement direction with coloured boxes (red: up, blue: down, orange: left, green: right). Zr/Hf (inactive in this mode) has been removed for clarity.}
    \label{fig:PnamModes}
\end{figure}

Looking now at the $Pbam$ dispersions (Figure \ref{fig:PbamPZOandPHO_phon}), we see that PHO and PZO retain closely related dynamical behaviour. One particular feature the eye is drawn towards is the instability of an optical branch in the vicinity of the Z-point. Since $Pbam$ is the purported low temperature ground state, this is surprising. The low temperature ground state (or more precisely, the 0K ground state) should have \textit{no unstable modes}. Exactly at the Z-point, this mode has irreducible representation (irrep) Z$_4^+$ for PZO and PHO with wavenumbers $26.15 i$ cm$^{-1}$ and $24.27 i$ cm$^{-1}$ respectively. By convention, mode irreps are usually given as a decomposition of the $Pm\bar{3}m$ phase, so, we unfold the single Z$_4^+$ irrep (of $Pbam$) to five irreps: T$_4$, T$_2$, $\Lambda_1$, $\Lambda_3$ and $\Delta_5$ where $\mathbf{q}_{\text{T}} = (\frac{1}{2}, \frac{1}{2}, \frac{1}{4})$, $\mathbf{q}_{\Lambda} = (\frac{1}{4}, \frac{1}{4}, \frac{1}{4})$ and $\mathbf{q}_{\Delta} = (0, \frac{1}{4}, 0)$. We point out that these distortions can all be located in the imaginary space of the $Pm\bar{3}m$ dispersions (Figure \ref{fig:Pm3mPZOandPHO_phon}) and were recorded in \cite{Baker2019}. The distortion patterns (though exaggerated) for T$_4$ and $\Lambda_3$ are shown in Figure \ref{fig:PnamModes}. We do not discuss the character of the remaining modes here since they only appear at a minuscule amplitude. T$_4$ (Figure \hyperref[fig:PnamModes]{3a}) is a long wavelength antiferrodistortive mode featuring octahedral rotations about the $c$ axis. It is periodic over four perovskite units with two octahedra rotating clockwise and two anticlockwise (a $++--$ pattern). This mode is reminiscent of the \textit{super-tilting} pattern observed in NaNbO$_3$ \cite{Chen1988} and AgNbO$_3$ \cite{Yashima2011}. $\Lambda_3$ (Figure \hyperref[fig:PnamModes]{3b}) is a Pb-O antipolar mode. For the Pb displacements, it can be described as having a `two-up, two-down' pattern in one PbO plane then a `two-left, two-right' pattern in the next PbO plane. Within these planes, O moves \textit{antiparallel} to the Pb displacements. Within the ZrO$_2$/HfO$_2$ planes, O moves in a sinusoidal wave pattern with a period of four O sites. This pattern is reflected (about the Pb-O plane) in the next ZrO$_2$/HfO$_2$ plane. \par

We now introduce the eigendisplacements associated with these new irreps into the $Pbam$ structure, breaking the symmetry and pushing the crystal into a new energy minimum. After relaxing the ionic positions, we arrive at an 80-atom $Pnam$ phase; a Klassengleiche maximal subgroup of $Pbam$ with $c$-axis doubling. This structure is lower in energy than $Pbam$ and described by \textit{eleven} distinct irreps. This is the sum of the five mentioned in the previous paragraph and the six pre-existing in the $Pbam$ structure \cite{iniguez2014}. To corroborate this energy lowering and to ensure this new phase is not a mere artefact of the LDA, we perform the same procedure with the PBESol \cite{Perdew2008} and SCAN \cite{Sun2015} functionals\footnote[2]{PBESol calculations were performed using norm-conserving pseudopotentials generated by the ONCVPSP code \cite{Hamann2013} (v0.3) using input from the PseudoDojo library \cite{vanSetten2018}. A 1088.46 eV plane wave cutoff was used. SCAN calculations were performed with \texttt{VASP} \cite{Kresse1996, Kresse1996I} (\texttt{v5.4.4}) using PAWs (Zr sv 04Jan2005, Hf pv 06Sep2000, O 08Apr2002, Pb d 06Sep2000) \cite{Blochl1994} and a 700 eV plane wave cutoff. We also remark that the lower energy of $Pnam$ compared with $Pbam$ is reproducible with LDA and PBESol calculations with \texttt{VASP} as well as with Wu-Cohen functional \cite{Wu2006} calculations with \texttt{SIESTA} \cite{Soler2002} (\texttt{v4.0}).}. The relative stabilities are shown in Table \ref{tab:relativestabilityPnam} and the resulting $Pnam$ and $Pbam$ crystal structures are given in Tables
\ref{tab:PnamCrystalstruct} and \ref{tab:PbamCrystalStruct} respectively. In anticipation of small energy differences, the results given in Tables \ref{tab:relativestabilityPnam}-\ref{tab:PbamCrystalStruct} are calculated with denser MP $\mathbf{k}$-point grids ($7 \times 5 \times 5$ for $Pbam$, $7 \times 5 \times 3$ for $Pnam$ and $8 \times 8 \times 8$ for $Pm\bar{3}m$, all $\Gamma$-centered). To quantify the strength of each distortion, we calculate the primitive cell normalised mode amplitude $A_p$ for each irrep which we display in Table \ref{tab:PnamModeDecomp}. $A_p$ is calculated by assigning atomic displacements (by symmetry) to an irrep and measuring the fractional displacements relative to the parent structure. We then normalise by a factor of $\sqrt{V_p/V_s}$ for primitive/supercell cell volumes $V_p/V_s$. $A_p$ is then the root sum squared (RSS) of each of these displacements comprising the irrep. This is the format popularized by the \texttt{ISODISTORT} package \cite{Campbell2006}. 

Table \ref{tab:relativestabilityPnam} shows that the new $Pnam$ phase is lower in energy than $Pbam$ for all three functionals used. $Pnam$ is more stable by $\sim$ 1 meV/FU for most cases, but, this narrows to $\approx$ 0.2 meV/FU for the SCAN functional. The majority of this energy lowering comes from the condensation of the T$_4$ and $\Lambda_3$ modes which appear with an amplitude similar to the $S_4$ mode of $Pbam$ in the LDA-PW and PBESol calculations. For SCAN, the amplitudes of these modes are degraded, explaining the narrowing of the energy difference between $Pbam$ and $Pnam$ for this functional. This effect is particularly apparent for $Pnam$ PHO where the amplitudes of these two modes are $\approx6 \times$ smaller compared with LDA-PW and PBESol. Even more interesting, the RSS of $Pnam$ PHO for the SCAN functional is \textit{lower} than $Pbam$; despite the introduction of five new modes, the total distortion decreases. This is the result of the reduced amplitude $R_4^+$ mode which competes with one of the new modes. It also worth mentioning that Tables \ref{tab:PnamCrystalstruct} and \ref{tab:PbamCrystalStruct} show that the lattice parameters (per ABO$_3$ unit) of the two models are almost unchanged ($\sim 10^{-3} \text{\AA}$ difference). It is therefore unlikely that we can experimentally distinguish between the two models by this simple comparison. \par

There are three possibilities we can conceive for the origin of this new $Pnam$ phase which should all be considered. Firstly, it could be that what was observed as $Pbam$ in experiment was $Pnam$ all along. This could be forgiven since distinguishing between the two models in any given measurement could be difficult, especially without prior knowledge of the $Pnam$ model. The new distortions are small in amplitude, so any measurement would likely have to be performed with high resolution equipment at cryogenic temperatures while taking great care to account for the possible presence complex twin domains \cite{Scott1972, Fesenko1976, Glazer1993}.
We remark that if the low temperature ground state is $Pnam$, it would be unsurprising seeing as we know the \textit{vast majority} of perovskites condense this symmetry at low temperatures \cite{Lufaso2001, Benedek2013} (although we reiterate that the 80-atom AFE $Pnam$ struture is different to the common perovskite prototype $Pnma$ structure). The second origin we have conceived is that in some region below the measured AFE phase transition temperature, the crystal is $Pbam$, but, at some point before 0K there is a second transition to $Pnam$, previously undetected due to its small magnitude. This origin, however, is unlikely. Using dielectric loss measurements of single crystal PZO, we observe no dielectric anomalies from room temperature to 10K strongly suggesting there is no such transition (unless of course the transition exists below 10K). While the presence of such an anomaly would indicate the presence of a phase transition, it does \textit{not} present the symmetry of the new phase. It is perhaps only careful neutron diffraction/scattering experiments on single crystal samples which would allow proper determination of $Pnam$ over $Pbam$. Confirmation would arise from the observation of $\Lambda$, $\Delta$ or T-point reflections. In particular, $\Lambda$-point reflections \textit{could} be characterised with X-rays as we expect a strong Pb character in the distortion. The third origin is that the new distortions are artefacts of the theoretical approach. At the level of DFT, we rely on the available functionals used to approximate the electronic exchange \& correlation interactions. While we have tried to minimize the possibility that these new distortions appear only with particular functionals (by using three at consecutively higher rungs which all predict $Pnam$ as a more stable phase than $Pbam$), we cannot explicitly rule this out. In addition, it is conceivable that while $Pnam$ is indeed the proper ground state atomic configuration, quantum and thermal fluctuations of nuclei are sufficient to suppress the condensation of some phonon modes and thus force the thermodynamical equilibrium atomic positions towards $Pbam$ at finite temperatures and pressures. Unfortunately, addressing the finite temperature properties of anharmonic crystals with unit cells as large as 80 atoms currently exceeds the computational tractability of \textit{ab initio} simulations. \par

While this $Pnam$ phase appears exotic, it is more common than one might think. Similar 80-atom AFE $Pnam$ phases are known to be metastable in BiFeO$_3$ (BFO) \cite{Prosandeev2014} and have recently been experimentally stabilized under the correct electrostatic boundary conditions \cite{Mundy2018}. They are also known to appear in a whole host of BFO-based solid solutions, including (Bi, La)FeO$_3$ \cite{Rusakov2011}, BiFe$_{0.75}$Mn$_{0.25}$O$_3$ \cite{Belik2011}, (Bi, Nd)FeO$_3$ \cite{Karimi2009, Levin2011} and BiFe$_{0.5}$Sc$_{0.5}$O$_3$ \cite{Prosandeev2014, Khalyavin2014}. We note that in all of these cases, the magnitude of the distortions defining $Pnam$ over $Pbam$ are \textit{much} stronger than what we find for PZO and PHO (for example, the energy difference between $Pbam$ and $Pnam$ is found to be 47 meV/FU in BFO \cite{Mundy2018} while this difference is only $\sim 1$ meV/FU in this work), making their experimental identification easier. It is perhaps not so surprising that there is a parity between the low energy AFE polymorphs of BFO and PZO/PHO; their polar distortions share a similar chemical origin. That is, they are thought to be driven by the polarizability of the s-orbital lone pair of Bi$^{3+}$ or Pb$^{2+}$ \cite{Seshadri2001, Hill1999, Cohen1992} distinguishing them from BaTiO$_3$ where polar distortions are driven by Ti 3d-O 2p hybridization \cite{Cohen1992}. \par

To summarize, we have shown that the phonon dispersions of the purported $Pbam$ AFE ground state of PZO and PHO are dynamically unstable. The eigendisplacements of the instability describe an 80-atom $Pnam$ phase slightly lower in energy than $Pbam$. Given that AFE phases of this type seem to be ubiquitous in BFO and BFO-based materials and it now appears (in DFT, at least) in the archetypal antiferroelectrics, the question must be asked: are 80-atom $Pnam$ phases the most common AFE arrangements in the perovskite oxides? While we cannot currently unequivocally declare that $Pnam$ is the true low temperature ground state structure of PZO and PHO, from the perspective of DFT, it is clear that $Pbam$ isn't. The experimental verification (or invalidation) of $Pnam$ symmetry would surely be challenging but as we have demonstrated, it is now time to take another look.

\section*{Acknowledgements}

J. S. Baker and D. R. Bowler are grateful for computational support from the UK Materials and Molecular Modelling Hub, which is partially funded by EPSRC (EP/P020194), for which access was obtained via the UKCP consortium and funded by EPSRC Grant Ref. No. EP/P022561/1. J. S. Baker and D. R. Bowler also acknowledge that this work used the ARCHER UK National Supercomputing Service funded by the UKCP consortium EPSRC Grant Ref. No. EP/P022561/1. J. K. Shenton acknowledges funding from the European Research Council (ERC) under the European Union’s Horizon 2020 research and innovation programme Grant agreement No. 810451. Computational resources were provided by ETH Zurich. We thank K. Roleder for supplying a single crystal PZO sample. We also acknowledge fruitful discussions with B. Grosso, N. Spaldin and N. Zhang. 

\nocite{*}

\bibliography{PZO_PHO.bib}

\newpage

\import{}{EnergyDiff.tex}

\import{}{PnamPositions.tex}

\import{}{PbamPositions.tex}

\import{}{ModeDecompPnam.tex}

\end{document}

%% file: EnergyDiff.tex
\begin{table}[hp]
\centering
\vspace{9.5cm}
\caption{The relative stability $\Delta E$ (in meV/FU) of the $Pbam$ and $Pnam$ phases compared to cubic $Pm\bar{3}m$ for PZO and PHO. $\Delta E = E(Pbam/Pnam) - E(Pm\bar{3}m)$.}
\medskip \medskip 
\label{tab:relativestabilityPnam}
\begin{tabular}{@{}cccc@{}}
\toprule \toprule
           & LDA-PW   & PBESol   & SCAN      \\ \midrule
$Pbam$ PZO & -310.744 & -262.240 & -258.943  \\
$Pnam$ PZO & -311.878 & -263.096 & -259.138  \\
$Pbam$ PHO & -95.309  & -175.014 & -205.946 \\
$Pnam$ PHO & -96.026  & -175.426 & -206.186  \\ \bottomrule \bottomrule
\end{tabular}
\end{table}

%% file: PnamPositions.tex
\begin{table}[hp]
\centering
\vspace{1.5cm}
\caption{The (fractional) Wyckoff positions ($x$, $y$, $z$) and orthorhombic lattice parameters for 80 atom $Pnam$ PZO and PHO as calculated with the LDA-PW, PBESol and SCAN functionals.}
\medskip \medskip
\label{tab:PnamCrystalstruct}
\begin{tabular}{@{}cccc@{}}
\toprule \toprule
Site             & LDA-PW                    & PBESol                     & SCAN                      \\ \midrule
\multicolumn{4}{c}{$Pnam$ PZO}                                                                        \\ \midrule
Pb 8d            & (0.7946, 0.8764, -0.0057) & (0.7963, 0.8757, -0.0061)  & (0.8044, 0.8760, -0.0035) \\
Pb 4c            & (0.7865, 0.8699, 0.2500) & (0.7883, 0.8689, 0.2500)  & (0.7947, 0.8691, 0.2500) \\
Pb 4c            & (0.2150, 0.1258, 0.2500) & (0.2122, 0.1260, 0.2500)  & (0.2049, 0.1281, 0.2500) \\ 
Zr 8d            & (0.7549, 0.6248, 0.1252) & (0.7554, 0.6246, 0.1252)  & (0.7585, 0.6243, 0.1251) \\
Zr 8d            & (0.2576, 0.8760, 0.6245) & (0.2589, 0.87636, 0.6246) & (0.2607, 0.8763, 0.6247) \\
O 8d             & (0.5094  0.4965, 0.8994) & (0.5105, 0.4962, 0.8978)  & (0.5044, 0.4987, 0.8985) \\
O 8d             & (0.4538, 0.7314, 0.8645) & (0.4571, 0.7326, 0.8649)  & (0.4615, 0.7351, 0.8630) \\
O 8d             & (0.4780, 0.7434, 0.3534) & (0.4815, 0.7451, 0.3542)  & (0.4727, 0.7410, 0.3569) \\
O 8d             & (0.7246, 0.6582, 0.4990) & (0.7267, 0.6569, 0.4990)  & (0.7241, 0.6558, 0.4994) \\
O 8d             & (0.0140, 0.5067, 0.6150) & (0.0152, 0.5071, 0.6150)  & (0.0072, 0.5033, 0.6158) \\
O 4c             & (0.6968, 0.5870, 0.2500) & (0.7010, 0.5887, 0.2500)  & (0.6978, 0.5923, 0.2500) \\
O 4c             & (0.3025, 0.3995, 0.2500) & (0.2967, 0.3985, 0.2500)  & (0.3004, 0.4001, 0.2500) \\
$a$ ($\text{\AA}$) & 5.8065                 & 5.8671                    & 5.9015                    \\
$b$ ($\text{\AA}$) & 11.6707                & 11.7505                   & 11.8087                  \\
$c$ ($\text{\AA}$) & 16.2246                & 16.3833                   & 16.4237                    \\ \midrule
\multicolumn{4}{c}{$Pnam$ PHO}                                                                        \\ \midrule
Pb 8d            & (0.7883, 0.8763, -0.0049) & (0.7895, 0.8758, -0.0049)  & (0.7998, 0.8746, -0.0010) \\
Pb 4c            & (0.7835, 0.8698, 0.2500) & (0.7846, 0.8692, 0.2500)  & (0.7920, 0.8695, 0.2500) \\
Pb 4c            & (0.2183, 0.1263, 0.2500) & (0.2161, 0.1263, 0.2500)  & (0.2078, 0.1298, 0.2500) \\
Hf 8d            & (0.7548, 0.6247, 0.1253) & (0.7551, 0.6246, 0.1251)  & (0.7591, 0.6237, 0.1249) \\
Hf 8d            & (0.2556, 0.8759, 0.6245) & (0.2571, 0.8761, 0.6246)  & (0.2597, 0.8766, 0.6248) \\
O 8d             & (0.5086, 0.4963, 0.8969) & (0.5092, 0.4963, 0.8958)  & (0.5011, 0.4997, 0.8952) \\
O 8d             & (0.4599, 0.7336, 0.8649) & (0.4639, 0.7350, 0.8646)  & (0.4714, 0.7394, 0.8625) \\
O 8d             & (0.4851, 0.7455, 0.3549) & (0.4862, 0.7463, 0.3561)  & (0.4742, 0.7409, 0.3611) \\
O 8d             & (0.7298, 0.6556, 0.4991) & (0.7317, 0.6546, 0.4992)  & (0.7290, 0.6522, 0.4999) \\
O 8d             & (0.0118, 0.5061, 0.6150) & (0.0128, 0.5063, 0.6150)  & (0.0018, 0.5008, 0.6166) \\
O 4c             & (0.7059, 0.5896, 0.2500) & (0.7085, 0.5918, 0.2500)  & (0.7064, 0.5988, 0.2500) \\
O 4c             & (0.2945, 0.3981, 0.2500) & (0.2901, 0.3983, 0.2500)  & (0.2932, 0.3994, 0.2500) \\
$a$ ($\text{\AA}$) & 5.7566                 & 5.8147                    & 5.8333                    \\
$b$ ($\text{\AA}$) & 11.5587                & 11.6482                   & 11.6571                   \\
$c$ ($\text{\AA}$) & 16.1342                & 16.2886                   & 16.2713                 \\ \bottomrule \bottomrule
\end{tabular}%
\end{table}

%% file: PbamPositions.tex
\begin{table}[hp]
\centering
\vspace{3.5cm}
\caption{The (fractional) Wyckoff positions ($x$, $y$, $z$) and orthorhombic lattice vectors for 40 atom $Pbam$ PZO and PHO calculated with the LDA-PW, PBESol and SCAN functionals. We also include a comparison with 10K neutron diffraction data.}
\label{tab:PbamCrystalStruct}
\medskip \medskip
\begin{tabular}{@{}ccccc@{}}
\toprule \toprule
Site             & LDA-PW                   & PBESol                   & SCAN                     & Exp (10K) \cite{Fujishita2003} \cite{Madigout1999}                \\ \midrule
\multicolumn{5}{c}{$Pbam$ PZO}                                                                                               \\ \midrule
Pb 4g            & (0.7035, 0.8770, 0.0000) & (0.7017, 0.8764, 0.0000) & (0.6951, 0.8762, 0.0000) & (0.6991, 0.8772, 0.0000) \\
Pb 4h            & (0.2868, 0.1275, 0.5000) & (0.2893, 0.1282, 0.5000) & (0.2952, 0.1297, 0.5000) & (0.2944, 0.1294, 0.5000) \\
Zr 8i            & (0.2431, 0.8754, 0.2497) & (0.2420, 0.8756, 0.2497) & (0.2401, 0.8760, 0.2497) & (0.2414, 0.8752, 0.2486) \\
O 4e             & (0.0000, 0.0000, 0.7713) & (0.0000, 0.0000, 0.7714) & (0.0000, 0.0000, 0.7689) & (0.0000, 0.0000, 0.7707) \\
O 4f             & (0.0000, 0.5000, 0.7999) & (0.0000, 0.5000, 0.7969) & (0.0000, 0.5000, 0.7974) & (0.0000, 0.5000, 0.7974) \\
O 4h             & (0.6961, 0.0928, 0.5000) & (0.7010, 0.0941, 0.5000) & (0.6985, 0.0958, 0.5000) & (0.6989, 0.0956, 0.5000) \\
O 4g             & (0.7230, 0.1587, 0.0000) & (0.7250, 0.1575, 0.0000) & (0.7237, 0.1560, 0.0000) & (0.7244, 0.1560, 0.0000) \\
O 8i             & (0.2431, 0.8754, 0.2497) & (0.5316, 0.7616, 0.7188) & (0.5328, 0.7619, 0.7197) & (0.5317, 0.7378, 0.7202) \\
$a$ ($\text{\AA}$) & 5.8098                   & 5.8716                   & 5.9028                   & 5.8736                   \\

$b$ ($\text{\AA}$) & 11.6864                  & 11.7651                  & 11.8129                  & 11.7770                  \\
$c$ ($\text{\AA}$) & 8.0993                   & 8.1776                   & 8.2078                   & 8.1909                   \\ \midrule
\multicolumn{5}{c}{$Pbam$ PHO}                                                                                               \\ \midrule
Pb 4g            & (0.7101, 0.8769, 0.0000) & (0.7092, 0.8764, 0.0000) & (0.7003, 0.8748, 0.0000) & (0.7114, 0.8768, 0.0000) \\
Pb 4h            & (0.2840, 0.1281, 0.5000) & (0.2855, 0.1285, 0.5000) & (0.2920, 0.1300, 0.5000) & (0.2928, 0.1298, 0.5000) \\
Hf 8i            & (0.2442, 0.8754, 0.2497) & (0.2434, 0.8756, 0.2497) & (0.2406, 0.8764, 0.2498) & (0.2421, 0.8745, 0.2455) \\
O 4e             & (0.0000, 0.0000, 0.7714) & (0.0000, 0.0000, 0.7714) & (0.0000, 0.0000, 0.7669) & (0.0000, 0.0000, 0.7650) \\
O 4f             & (0.0000, 0.5000, 0.7951) & (0.0000, 0.5000, 0.7927) & (0.0000, 0.5000, 0.7905) & (0.0000, 0.5000, 0.7932) \\
O 4h             & (0.7049, 0.0950, 0.5000) & (0.7083, 0.0959, 0.5000) & (0.7065, 0.0995, 0.5000) & (0.6996, 0.0983, 0.5000) \\
O 4g             & (0.7284, 0.1560, 0.0000) & (0.7303, 0.1551, 0.0000) & (0.7290, 0.1522, 0.0000)   & (0.7329, 0.1561, 0.0000) \\
O 8i             & (0.5282, 0.7609, 0.7195) & (0.2434, 0.8756, 0.2497) & (0.5271, 0.7599, 0.7234)   & (0.5280, 0.7410, 0.7190) \\
$a$ ($\text{\AA}$) & 5.7585                   & 5.8181                   & 5.8330                   & 5.8404                   \\
$b$ ($\text{\AA}$) & 11.5690                  & 11.6580                  & 11.6575                  & 11.7057                  \\
$c$ ($\text{\AA}$) & 8.0581                   & 8.1344                   & 8.13517                  & 8.1751                   \\ \bottomrule \bottomrule
\end{tabular}%
\end{table}

%% file: ModeDecompPnam.tex
\begin{table}[hp]
\centering
\vspace{6.5cm}
\caption{The total decomposed mode amplitudes $A_p$ (described in the text) for each irrep using the $Pm\bar{3}m$ phase as the parent and the $Pbam$/$Pnam$ phase as the daughter. Data is presented in the format  ``LDA-PW PBESol SCAN". RSS = $\sqrt{\sum_i A_{p, i}^2}$.}
\medskip \medskip
\label{tab:PnamModeDecomp}
\begin{tabular}{@{}ccccc@{}}
\toprule \toprule
Mode        & $Pbam$ PZO           & $Pbam$ PHO           & $Pnam$ PZO           & $Pnam$ PHO           \\ \midrule
R$_4^+$     & 0.5504 0.5332 0.5138 & 0.5082 0.4969 0.4433 & 0.5380 0.5197 0.5088 & 0.4964 0.4841 0.4418 \\
$\Sigma_2$ & 0.4233 0.4093 0.4462 & 0.3541 0.3416 0.3832 & 0.4098 0.3942 0.4440 & 0.3415 0.3308 0.3838 \\
S$_4$       & 0.1412 0.1296 0.1424 & 0.1191 0.1094 0.1192 & 0.1443 0.1301 0.1427 & 0.1206 0.1114 0.1186 \\
M$_5^-$     & 0.0096 0.0156 0.0257 & 0.0103 0.0155 0.0354 & 0.0159 0.0230 0.0265 & 0.0153 0.0197 0.0380 \\
R$_5^+$     & 0.0277 0.0277 0.0376 & 0.0305 0.0302 0.0301 & 0.0267 0.0260 0.0349 & 0.0276 0.0268 0.0294 \\
X$_3^-$     & 0.0191 0.0173 0.0169 & 0.0159 0.0137 0.0127 & 0.0135 0.0107 0.0153 & 0.0101 0.0092 0.0127 \\
T$_4$       & -                    & -                    & 0.1337 0.1420 0.0641 & 0.1283 0.1261 0.0160 \\
$\Lambda_3$ & -                    & -                    & 0.1275 0.1278 0.0735 & 0.1116 0.1007 0.0181 \\
$\Lambda_1$ & -                    & -                    & 0.0264 0.0263 0.0155 & 0.0227 0.0201 0.0046 \\
T$_2$       & -                    & -                    & 0.0046 0.0046 0.0028 & 0.0030 0.0026 0.0007 \\
$\Delta_5$  & -                    & -                    & 0.0113 0.0132 0.0092 & 0.0136 0.0086 0.0024 \\
RSS         & 0.7094 0.6855 0.6969 & 0.6318 0.6139 0.5999 & 0.7172 0.6936 0.6988 & 0.6390 0.6196 0.5997 \\ \bottomrule \bottomrule
\end{tabular}%
\end{table}